\documentclass[12pt,preprint]{aastex}

\shorttitle{Collapsed Cores in Globular Clusters}
\shortauthors{Djorgovski et al.}

\begin{document}


\title{Chandra Observation of the Interaction between the Hot Plasma
    Nebula RCW89 and the Pulsar Jet of PSR B1509--58}


\author{Y. Yatsu\altaffilmark{1}, J. Kataoka\altaffilmark{1},
N. Kawai\altaffilmark{1}, and T. Kotani\altaffilmark{1}}
\affil{Tokyo Institute of Technology, 2-12-1 Ohokayama, Meguro, Tokyo
Japan 152-8551}
\email{yatsu@hp.phys.titech.ac.jp}
\author{K. Tamura\altaffilmark{2}}
\affil{Nagoya University}
\and
\author{W. Brinkmann\altaffilmark{3}}
\affil{Max-Planck-Institut f\"{u}r Extraterrestrische Physik \\
   Postfach 1603, 85740 Garching, Germany }

\begin{abstract}
 We present a {\itshape Chandra} observation of the H {\rm \footnotesize
II} region RCW89.  The nebula lies on 10' north from the central pulsar
PSR B1509--58, and it has been suggested that the nebula is irradiated
by the pulsar jet.  We performed a spectral analysis of the seven
brightest emitting regions aligned in a ``horse-shoe'' shape, and found
that the temperature of the knots increases along the ``horse-shoe'' in
the clockwise direction, while, in contrast, the ionization parameter
$n_e t$ decreases.  This strongly supports a picture of energy transfer
via the precessing pulsar jet.  We examined the energy budget assuming
that RCW89 is powered by the pulsar jet, and confirmed that the
pulsar rotational energy loss is sufficient to drive the nebula.  The
rate of energy injection into RCW89 by the jet was estimated from the
synchrotron radiation flux.  We obtained a heating time-scale of 1400
yr, which is consistent with the pulsar characteristic age of 1700 yr.
To explain the temperature gradient, we discuss the cooling process for
plasma clouds in RCW89.  We argue that the plasma clouds can be cooled
down by the adiabatic expansion within 70 yr, and form the temperature
gradient reflecting the sequential heating by the precessing pulsar
jet.  We also determined the velocities of the individual plasma clouds
by spectral fitting.  The plasma clouds in RCW89 are moving
away at $240 \sim 860$ km s$^{-1}$, which constrains the inclination angle
of the pulsar spin axis $i > 50^{\circ}$ and the expanding velocity of
the shell as $v_{\mbox{\footnotesize shell}} > 1100$ km s$^{-1}$.
\end{abstract}




\section{Introduction}
Recent high-resolution X-ray imaging of young energetic pulsars have
revealed the ubiquitous presence of ``jets'' or collimated outflows
along the symmetry axis, as seen in the Crab pulsar (Brinkmann,
Aschenbach, and Langmeier, 1985; Hester et al. 1995; Weisskopf et
al. 2000), the Vela pulsar (Helfand, Gotthelf, \& Halpern 2001; Pavlov
et al. 2002), and PSR B1509-58 (Trussoni et al. 1996; Gaensler et
al. 2002).  These jets are probably formed by the relativistic outflows
of pulsar magnetospheres, but their composition, energy contents,
collimation mechanism, and their relation to the ring- or disk-like
pulsar wind nebulae have not been understood.  Investigations of pulsar
jets allow us to understand the mechanism of particle acceleration and
interaction between the pulsar wind and the interstellar medium.
Particularly intriguing is the case with PSR B1509--58, which is
accompanied by an unidentified thermal nebula RCW89 on the terminus of
the pulsar jet.  Because of its location, RCW89 has been considered to
be a contact point of the jet and the surrounding SNR or the ambient
medium.  If it is indeed the case RCW89 could be a precious probe
for the mysterious pulsar jet.

 The supernova remnant (SNR) G320.4--01.2 (MSH15-52, Kes23) has
bilateral radio shells with a diameter of $\sim$ 30 arcmin (Caswell,
Milne \& Wellington 1981).  PSR B1509--58, a 150 ms X-ray, radio, and
gamma-ray pulsar, is located at the center of G320.4--01.2 (Seward \&
Harnden 1982; Manchester, Tuohy, \& D'Amico 1982).  From the spin
parameters, a characteristic age $\tau_c = 1700$ yr (Kaspi et al. 1993),
a spin-down luminosity $\dot{E}=1.8 \times 10^{37}$ erg s$^{-1}$, and a
surface magnetic field $B_p=1.5\times10^{13}$ G have been obtained,
making it one of the youngest, the most energetic, and the highest field
pulsars ever known.  The age of 6$-$20 kyr for the SNR has been derived
by assuming standard parameters for the ISM and for the supernova
explosion.  But this age is an order of magnitude larger than the
characteristic age of the pulsar.  Although the SNR has a complex
structure with multiple components, H {\rm \footnotesize I} absorption
measurements confirm that it is a single SNR at a distance of
5.2$\pm$1.4 kpc (Gaensler et al. 1999), suggesting that both the pulsar
and the SNR originate from the same supernova.

The H$\alpha$ emission nebula RCW89 was discovered by Rogers, Campbell,
\& Whiteoak (1960) and coincides with the brightest portion of
G320.4--01.2 in radio.  It also coincides with one (north nebula) of the
two clearly distinguishable X-ray nebulae observed by the {\itshape
Einstein} Observatory (Seward et al. 1983).  The X-ray images taken by
{\itshape ROSAT} show the distribution of the plasma clumps into a
``horse-shoe'' like shape (Trussoni et al. 1996; Brazier et al. 1997).
Two possible candidates have been proposed for the energy source of the
thermal nebula: the supernova blast wave, and the pulsar jet (Manchester
1987).  From the analysis of {\itshape ASCA} data, Tamura et al. (1996)
found a bridge of non-thermal emission connecting the pulsar to the
thermal emission.  They also found that in RCW89 region thermal emission
with prominent line features is dominant.  The analysis of the thermal
X-ray spectra showed that its energy content was consistent with being
powered by the pulsar through the jet.  They also argued that a less
collimated jet or the precession of the jet is required for heating the
spatially extended plasma.

Gaensler et al. (1999) reported detailed radio observations using the
Australia Telescope Compact Array (ATCA).  Due to the correspondence of
the radio synchrotron knots with the clumps in the X-ray image, they
propose that the pulsar is interacting with RCW89 through the jet, and
that the non-thermal knots are embedded in the diffuse nebula emitting
the thermal X-rays.  This picture was however questioned by the
{\itshape Chandra} data showing that all the knots in RCW89 have thermal
spectra (Gaensler et al.2002).  Thus the model involving thermal clumps
embedded in a diffuse synchrotron nebula (as proposed by Tamura et
al. 1996) seems more likely.  In this case the energy budget discussed
by Tamura et al. (1996) must be reconsidered; {\itshape Chandra} has
revealed that the energy content in the knots is by two orders of
magnitude smaller than that deduced from the {\itshape ASCA}
observations.

\section{Observation and Results}
G320.4--01.2 was observed on 2000 August 14 with the Advanced CCD
Imaging Spectrometer (ACIS) I-array aboard {\itshape Chandra}.  The net
exposure time was 19039 s after removing small intervals in which no
data were recorded (Gaensler et al. 2002).  A smoothed X-ray image in
the energy range 0.4$-$8.0 keV is shown in Figure \ref{X-ray image}.

PSR B1509--58, accompanied by the bright PWN, is at the center of the
FOV of the ACIS-I and the jet features are extended toward the
north-west and south-east direction.  RCW89 lies on the end point of the
north-west jet, exhibiting a complicated clumpy structure.  We extracted
photons from the brightest seven emitting regions (labeled as 1$-$7 in
clockwise direction in Fig. \ref{X-ray image}) and the jet region for
comparison, using the {\it psextract} script of CIAO 2.3 and carried out
a spectral analysis.  The background was taken from the area marked in
Figure \ref{X-ray image}.  Figure \ref{spectra2} shows the spectra of
the north jet and region A.  The spectrum of the north jet is well
fitted by a single power law function with a photon index of $\Gamma =
2.07\pm0.11$ and a hydrogen absorption column density of N$_{\rm H} =
(0.86\pm0.09)\times 10^{22}$ cm$^{-2}$ with a $\chi^2_{\rm red} = 0.981
$ for 70 d.o.f..  The absorption corrected flux density for the energy
range 0.5$-$5.0 keV is $(1.26\pm0.09)\times 10^{-12}$ ergs s$^{-1}$.

The energy spectra of RCW89 regions are quite complex and show strong
emission lines as well as a hard high energy tail, which has already
been found in the ASCA data (Tamura et al 1996).  For the analysis of
individual knots in RCW89, the photon statistics was not sufficient to
determine the column density $N_H$ and the photon index $\Gamma$ for
each knot independently.  To determine the spectral parameters for each
knot we first analyzed the spectrum of the total region A in Figure
\ref{X-ray image}, which contains all of the bright knots 2$-$7.  The
spectrum of the region ``A'' is shown in Figure \ref{spectra2}.  We
applied two component model consisting of a non-equilibrium ionization
model (``VNEI'' in XSPEC 11.3.0) for the thermal component and a power
law for the high energy tail.  In the NEI model, the metal abundances of
Ne, Mg, Si, and Fe, which all have remarkable line emission, were set to
be variable.  The best fit parameters are shown in Table 1.  The column
density and the photon index were obtained as $N_H = (1.18\pm 0.01)
\times 10^{22}$ cm$^{-2}$ and $\Gamma = 2.48 \pm 0.05$ with a $\chi_{\rm
red}^2 =1.727$ for 227 d.o.f.  The fit is rather poor which is mainly
due to large residuals in the energy range of 1.0$-$1.3 keV.  For the
individual spectral analysis we used fixed values of $N_H = 1.18\times
10^{22}$cm$^{-2}$ and $\Gamma=2.48$.  The results of the spectral
analysis for each knot are shown in Table 2.

\begin{figure}[h]
\begin{center}
\includegraphics[height=8.5cm,angle=-90]{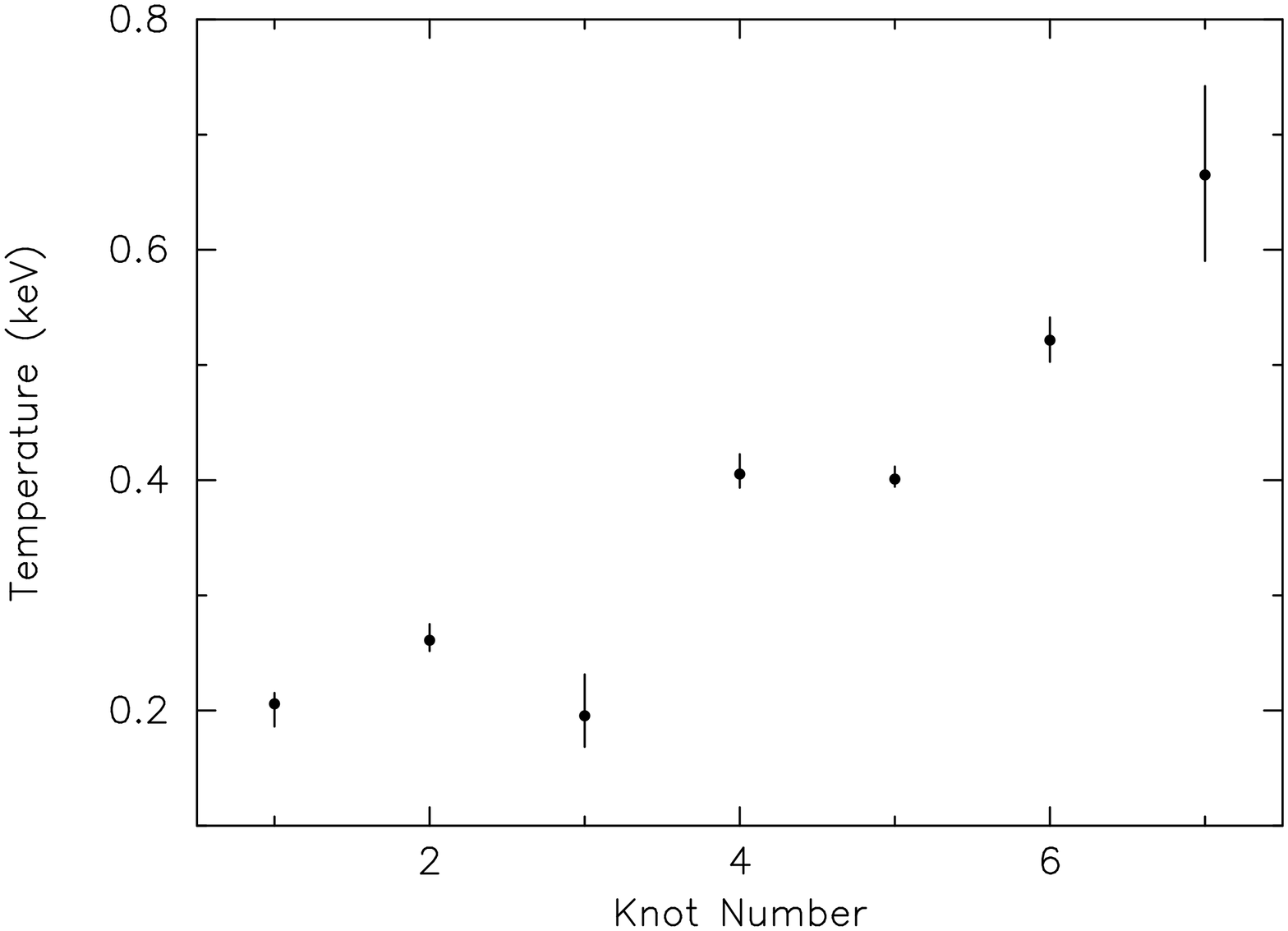}
\includegraphics[height=8.5cm,angle=-90]{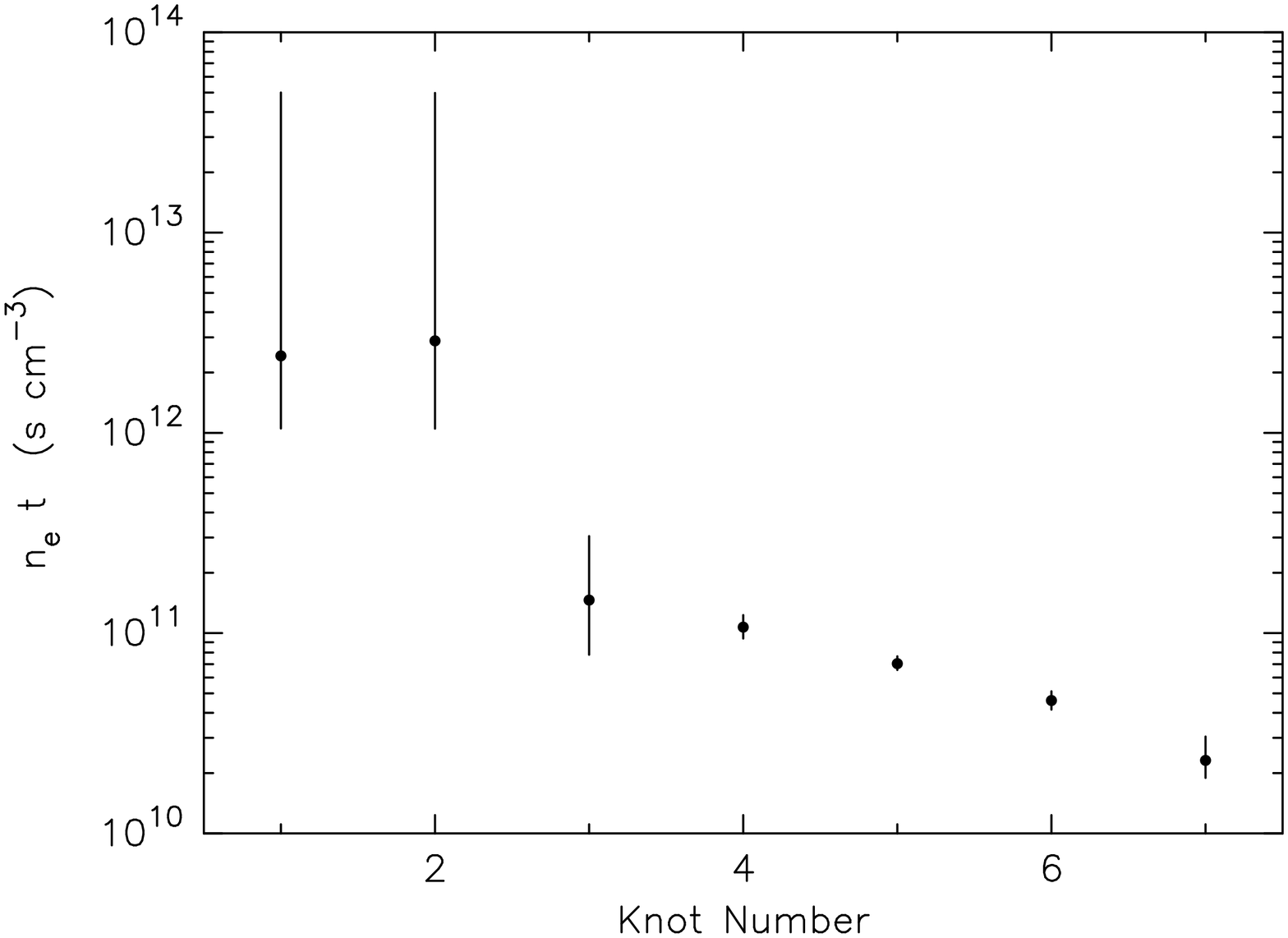} 
\caption{Spatial
distribution of the temperature (Left) and the ionization time scale
$n_e t$ cm$^{-3}$s (Right).  The plasma temperature increases along the
``horse-shoe'' in clockwise direction, while in contrast, $n_e t$
decreases.  Uncertainties are at 1 $\sigma$ confidence.}
\label{distribution}
\end{center}
\end{figure}

Figure \ref{distribution} shows the temperatures and ionization
parameters $n_e t$ from the individual fits to the knots of RCW89.  This
allows a crude estimate of the physical age of each cloud.  The
temperatures tend to increase along the ``horse-shoe'' in a clockwise
direction from knots 1 to 7, while the ionization parameter decreases.
This result implies that the knots on the north edge of RCW89 consist of
freshly heated hot plasma and are still in the state of non-equilibrium
ionization while the clouds in the southern regions are close to
ionization equilibrium.

\section{Discussion}
In the previous sections, we show clear evidence for variations of
plasma parameters (temperature and age) in bright knots in RCW89.  There
are two possibilities to account for the thermal X-ray emission from
RCW89, (1) shock heating due to the supernova blast wave and (2) heating
by the pulsar jet of PSR B1509--58.  To constrain the energy source of
RCW89, we evaluate the ages of the plasma clouds.  The {\itshape
Chandra} image yields the volumes for emitting plasma clouds, assuming
that the shapes of the clouds are ellipsoids with major and minor axes
of $l_a$ and $l_b$ in arcsec.  We obtain the number densities of the
electrons in the clouds as
\begin{equation}
n_e = 7.0 \times 10^{2} 
\left(l_a \; l_b^2 \right)^{-1/2}
\left(\frac{E\!M}{10^{56}\mbox{ cm}^{-3}}\right)^{1/2} 
\left(\frac{d_L}{5.2 \mbox{ kpc}}\right)^{-1/2}
\quad \mbox{cm}^{-3},
\end{equation}
where $E\!M$ is the emission measure for each cloud derived from the
spectral fits above.  Dividing the ionization parameter $n_e t$ by the
electron number density $n_e$ provides the ionization age $t_{\mbox{p}}$.
\begin{equation}
 t_{\mbox{p}} = 4.5 \times
\left(l_a \; l_b^2\right)^{1/2} 
\left(\frac{n_e t}{10^{11}\mbox{ s cm}^{-3}}\right) 
\left(\frac{E\!M}{10^{56}\mbox{ cm}^{-3}}\right)^{- 1/2} 
\left(\frac{d_L}{5.2 \; \mbox{ kpc}}\right)^{1/2}
 \quad \mbox{yr}.
\end{equation}
The resulting ages of the plasma clouds are shown in Table 3.  Their
ages are in the ranged of 5 $\sim$ 700 yr, which is clearly different from each
other.  Thus, it is not likely that all the knots of RCW89 were heated
simultaneously by the supernova blast-wave.

Next, we examine the second hypothesis (that the knots were heated by
the pulsar jet) based on the energy budget of the plasma clouds.  The
internal thermal energy stored in a plasma cloud is estimated as
\begin{equation}
E_{th} = 6.7 \times 10^{45} 
\left(l_a \; l_b^2 \right)^{1/2} 
\left( \frac{E\!M}{10^{56}\mbox{ cm}^{-3}}\right)^{1/2}
\left( \frac{kT}{1\mbox{ keV}}\right)
\left(\frac{d_L}{5.2 \; \mbox{ kpc}}\right)^{5/2}
\quad \mbox{ergs},
\end{equation}
which are shown as $E_{th}$ in Table 3.  The total energy in the seven
knots amounts to $2.6\times10^{47}\left(d_L/5.2\mbox{kpc}\right)^{5/2}$
ergs.  The heating time scale can be constrained by taking the
observable pulsar jet flow into account.  The non-thermal flux in the
north jet region represents the synchrotron emission from the
relativistic electrons.  From the spectral fit in Table 1, we therefore
infer that unabsorbed spectral luminosity at 1 keV is
$L_{\mbox{{\footnotesize 1 keV}}} = 2.1 \times 10^{33}$ ergs s$^{-1}$
keV$^{-1}$ for a distance of $d_L = 5.2$ kpc.  We assume the jet to be a
cylinder of length $l_j = 3\times 10^{18}$ cm and cross section $S_j =
2.0\times 10^{35}$ cm$^{2}$.  Assuming equipartition between electrons
and magnetic field, we obtain the minimum energy of electrons and
magnetic field in the jet,
\begin{equation}
E_{jet}=6.4\times 10^{35}(1+\mu)^{4/7}f^{3/7} \left(\cos i \right)^{-3/7}
 \quad \mbox{ergs},
\end{equation}
where $f$ is the filling factor, and $\mu$ is the ratio of ion to
electron energy.  $i$ is the inclination angle of the jet.  The obtained
energy $E_{jet}$ corresponds to a magnetic field $B_{jet}\sim 34\mu$G in
the north jet, which is consistent with $B \sim 25\mu$G in the south jet
evaluated by Gaensler et al. (2002).  The observed photon index for the
north jet is $\sim$2.0, while the south jet in the vicinity of the
pulsar has a photon index of 1.6.  This result implies that the
synchrotron electrons emitting in the observed energy band had been
cooled before the jet reached the terminal.  In a magnetic field $B\sim
34\mu$G, the life time of synchrotron electrons which emit 1 keV photons
is $\sim 50$ yr.  Assuming that the electrons are accelerated in the
vicinity of the pulsar, the life time of the electron constrain the
velocity of the north jet,
\begin{equation}
 v_{jet} \leq 0.5 c \left(\frac{l}{7.5 \mbox{pc}}\right)
\left(\frac{t_{cool}}{50\mbox{ yr}}\right)^{-1}
\end{equation}
where $l$ is the distance between the ``jet (north)'' region marked in Figure
\ref{X-ray image} and the central pulsar.  

For a bulk velocity of the jet $v_{\mbox{\tiny BLK}} \sim 0.5c$,
analogous to that of the Crab nebula(Hester et al. 2002), the power
injected into RCW89 amounts to
\begin{equation}
\dot{ E}_{jet}= 3.2 \times 10^{35}(1+\mu)^{4/7}f^{3/7}
\left( \cos i\right)^{4/7}
\left( \frac{v_{\mbox{\tiny BLK}}}{0.5c} \right) \quad \mbox{ergs s$^{-1}$}.
\end{equation}
This corresponds to 1.7\% of the pulsar's total spin-down luminosity.
If the ratio at which the pulsar supplies energy to the north jet had
not changed, we can evaluate the initial rotation period as 35 ms for a
pulsar's moment of inertia $10^{45}$ g cm$^2$.  With a constant braking
index of $n = 2.8$, a pulse period of $P=150$ msec, and its first
derivative $\dot{P}=1.5\times 10^{-12}$ s s$^{-1}$ (Kaspi et al. 1993),
the required time to heat RCW89 is then 1400 yr.  The ages of oldest
plasma estimated from the ionization time scale and the heating time
scale required by the energy budget are both of the order of $\sim 10^3$
yr, in a good agreement with the pulsar's characteristic age of 1700 yr.
Thus our results favors the scenario of heating by the pulsar jet rather
than the simultaneous shock heating by the blast wave.

With the obtained magnetic field, we can also estimate the luminosity in
TeV range.  In the magnetic field of 34 $\mu$G, the Lorentz factor of
the electrons which emit synchrotron photons at 1 keV is $\sim 2.3
\times 10^8$.  Such ultra-relativistic electrons can up scatter CMB
photons up to $\sim$ 1 TeV gamma-ray.  The ratio of the inverse Compton
luminosity to the synchrotron luminosity can be described as
$L_{IC}/L_{sync} \approx u_{CMB}/u_{mag}$, where $u_{CMB}$ and $u_{mag}$
are the energy densities of the cosmic microwave background and that of
the magnetic field, respectively.  Here, we must compare the
synchrotron/inverse Compton luminosity correspond to the same population
of relativistic electrons.  The observed synchrotron luminosity in
0.5-5.0 keV is $L_{sync} \sim 4.0 \times 10^{33}$ ergs s$^{-1}$.  We
infer a luminosity of the inverse Compton component in the north jet
$L_{IC} \sim 3.0\times 10^{31}$ ergs s$^{-1}$ in 1.3-13 TeV which
corresponds to the energy of CMB photons scattered by the synchrotron
electrons emitting in 0.5-5.0 keV.  Since the expected energy flux of
$1.0\times 10^{-15}$ ergs s$^{-1}$ cm$^{-2}$ is lower than the detection
limit of {\itshape HESS}, we cannot expect to detect the TeV gamma-rays
from the north jet.

The observed temperature of the elder plasma clouds are cooler.  This
result implies that an effective cooling process was working and have
formed the observed temperature gradient.  To investigate how the
temperature gradient was formed we consider the cooling process of the
plasma clouds.

First we examine the radiative cooling.  The dominant emission processes
from RCW89 are bremsstrahlung and line emission.  However, the radiation
fluxes observed by {\itshape Chandra} are too small to explain the
temperature gradient.  We find that the cooling time scale of the knots
due to optically thin thermal radiation is longer than $\sim$ 4000 yr
which is much longer than the pulsar age, and cannot explain the
temperature gradient.

Another plausible process is an adiabatic cooling.  We discuss two
possible cases of adiabatic expansion.  In the first case (``{\itshape
pure expansion}''), the heated gas volume expands without mixing with
the ambient matter.  In the second case (``{\itshape Sedov}''), the hot
gas expands like in the {\itshape Sedov} phase of an SNR, where the
expanding shock front mixes and heats the ambient matter.

 In the ``{\itshape pure}'' adiabatic expansion, the temperature
{\itshape T} and the volume {\itshape V} follow the relation $T
V^{\gamma - 1} = \mbox{const}$, ($\gamma$ is the specific heat ratio;
$\gamma = 5/3$ for a monoatomic gas).  We consider, for example, the
case of the plasma cloud labeled ``7'' in Figure \ref{X-ray image} with
a radius of $r_0 = 2 \times 10^{17}$ cm and a temperature of $T_0 = 0.6$
keV.  From the $V-T$ relation, a radius of $r = 2.8 \sim 10^{17}$ cm is
expected when the knot cools down to a temperature of 0.3 keV.  Assuming
that the expansion velocity equals to the sound velocity in the plasma,
$c_s = \left(\frac{\gamma p}{\rho} \right)^{1/2}$ and the
thermal pressure of $p=2n_e k T$, the expansion velocity can be
expressed as a function of radius $r$ as
\begin{equation}
\frac{dr}{dt}=\left(\frac{2\gamma k T_0 r_0^2}{m_p}\right)^{1/2} \frac{1}{r}.
\end{equation}
Integration yields the time required to expand from $r_0$ to $r$ as
\begin{equation}
 t=\sqrt{\frac{m_p}{8 \gamma k T_0 r_0^2}}(r^2-r_0^2)
\end{equation}
From this equation, the cooling time is estimated as $t_{cool} \sim 70$
yr.  In this calculation we have assumed that the density of the plasma
cloud is much higher than that of the ambient ISM, which thus does not
affect the expansion velocity.

Next, we examine the ``{\itshape Sedov}'' case.  The radio observations
showed that RCW89 lies on the north shell of the SNR (Gaensler et
al. 1999, 2002).  Therefore, the jet may heat a small volume of the
north shell to form a hot plasma which expands into the ambient matter.
In the Sedov model the injected energy $E_0$ is divided into internal
energy and into kinetic energy as
\begin{equation}
 E_0 = \frac{4}{3}\pi r^3\left(\frac{p}{\gamma -1} + \frac{\rho v^2}{2}\right),
\end{equation}
where $p$ is the pressure, $\rho$ is mass density, and $v$ is the
expansion velocity.  We crudely approximate the velocity as
$v \sim c_s$, which yields
\begin{equation}
 v = \left(\frac{15}{28\pi}\right)^{1/2} \left(\frac{E_0}{\rho r^3}\right)^{1/2},
\label{v-function}
\end{equation} 
where $E_0$ is the injected energy.  For a radius of $r=2 \times 10^{17}$
cm, velocity was estimated to be $v \sim 400$ km s$^{-1}$.  
The radius $r$ can be described as a function of time $t$ by integrating
the velocity
\begin{equation}
 r=1.0 \left(\frac{E_0}{\rho}\right)^{1/5} t^{2/5}.
\end{equation}
The relation between the temperature $kT$ and the radius $r$ is obtained
by substituting the equation of the state $p=2n_e kT$ into the
Eq. (\ref{v-function}) as

\begin{equation}
kT = \frac{9}{28\pi} \frac{E_0}{(n_p + n_e) r^3},
\end{equation}
where $k$ is Boltzmann constant and $n_p$ and $n_e$ are the number
densities of proton and electron, respectively.  Then we obtain the time
scale to cool from $kT_0$ to $kT$ as
\begin{equation}
 t_{cool}= 
\left( \frac{9}{56 \pi }\right)^{5/6} m_p^{1/2}
\left(\frac{E_0}{n_e}\right)^{1/3}
\left[
\left( \frac{1}{kT}\right)^{5/6} - \left(\frac{1}{kT_0}\right)^{5/6} 
\right].
\end{equation}
For knot-7 ($r_0= 2 \times 10^{17}$ cm, $T\sim 0.6$ keV, $n_e = 130
$cm$^{-3}$ ) the initially injected energy was estimated as $E_0 \sim
1.3 \times 10^{46} $ ergs.  The time required to cool from 0.6 keV to
0.3 keV is about $\sim$ 40 years.  

Both the ``{\itshape pure } expansion'' and ``{\itshape Sedov} model''
can provide sufficient cooling to form the temperature gradient within a
heating time scale.  At this moment it is hard to say which process is
more appropriate since we do not understand well enough the environment
surrounding the plasma clouds.  Moreover, the knots are embedded in a
diffuse synchrotron emission.  It may be therefore possible that these
plasma clouds are confined by the ambient magnetic field, and their
expansion may not be as free as we have assumed here in the discussion.

The differences in the temperatures and plasma ages of the knots
indicate that RCW89 was heated in sequence.  If these plasma clouds have
been heated by the pulsar jet, the direction of the jet would vary with
the period.  From the tendency of the temperature and the plasma age, we
infer that the interaction point between the jet and the ISM has shifted
along the ``horse-shoe'' in clockwise direction.  This implies the
``precession like'' motion of the jet.  Preceding discussions about the
energy budget and the cooling process indicate that the jet has made
only one precession since the birth of the pulsar within a time scale of
1000 yr.  

Considering the origin of the matter which forms the hot clouds, the
abundance of elements may provide a precious clue.  From the spectral
fitting we have obtained the metal abundances of the knots in RCW89.
The observed iron abundances tend to be smaller than the solar
abundance, while Ne and Mg are relatively rich as shown in Table 2.
This suggests that the plasma in RCW89 may originate from a type-II
supernova.  If RCW89 consists of the ejecta from the supernova
associated with the central pulsar, the ejecta must have traveled at
least $\sim 7.5$ pc from the SNR center.  In order to arrive at RCW89
within 1700 yr, the required velocity of the ejecta is more than
$\sim4000$ km s$^{-1}$, which is rather fast for the standard ISM
condition.  

\begin{figure}[h]
\includegraphics[height=7.5cm,angle=-90]{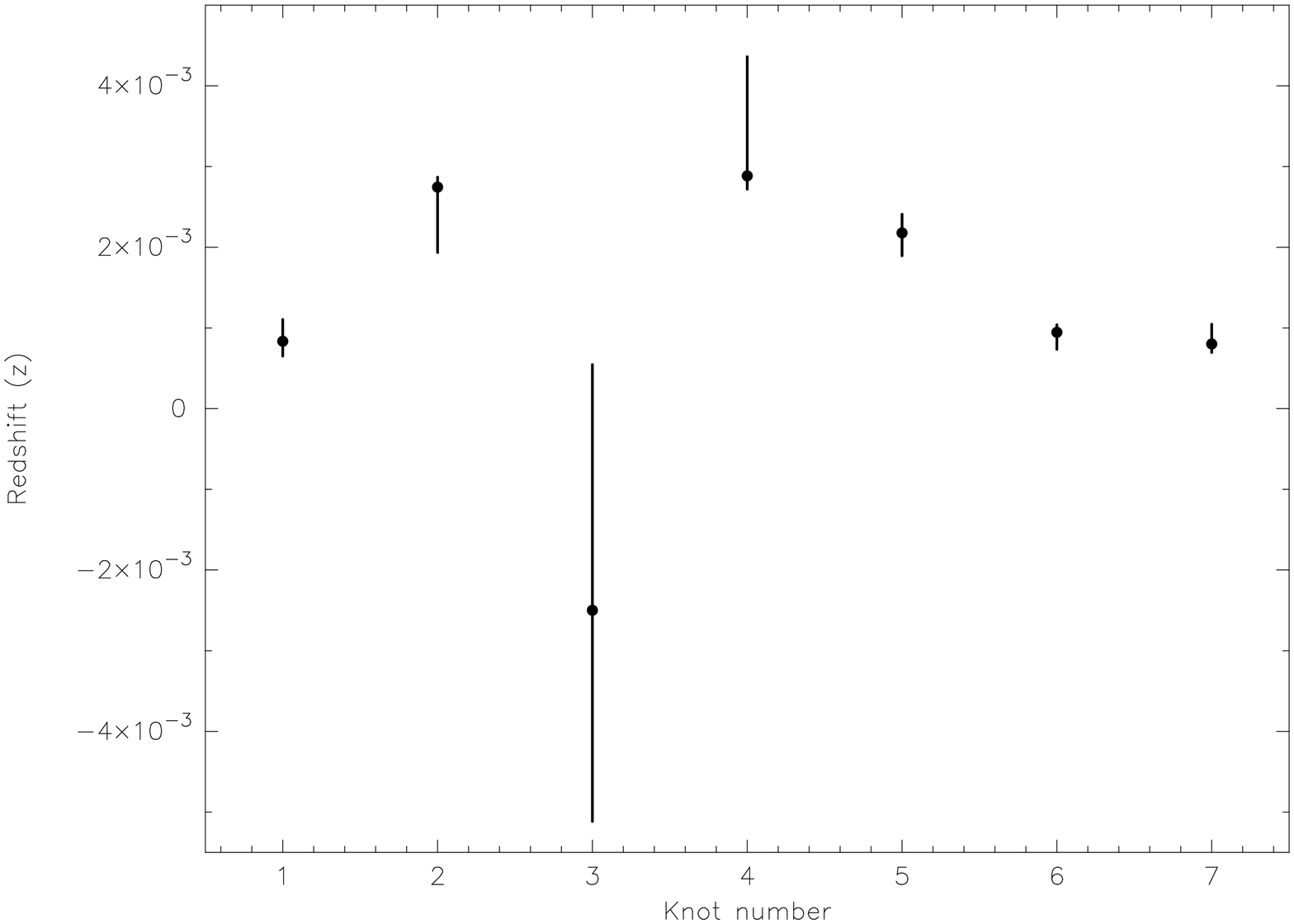}\hspace{1.0cm}
\includegraphics[height=7cm,angle=-90]{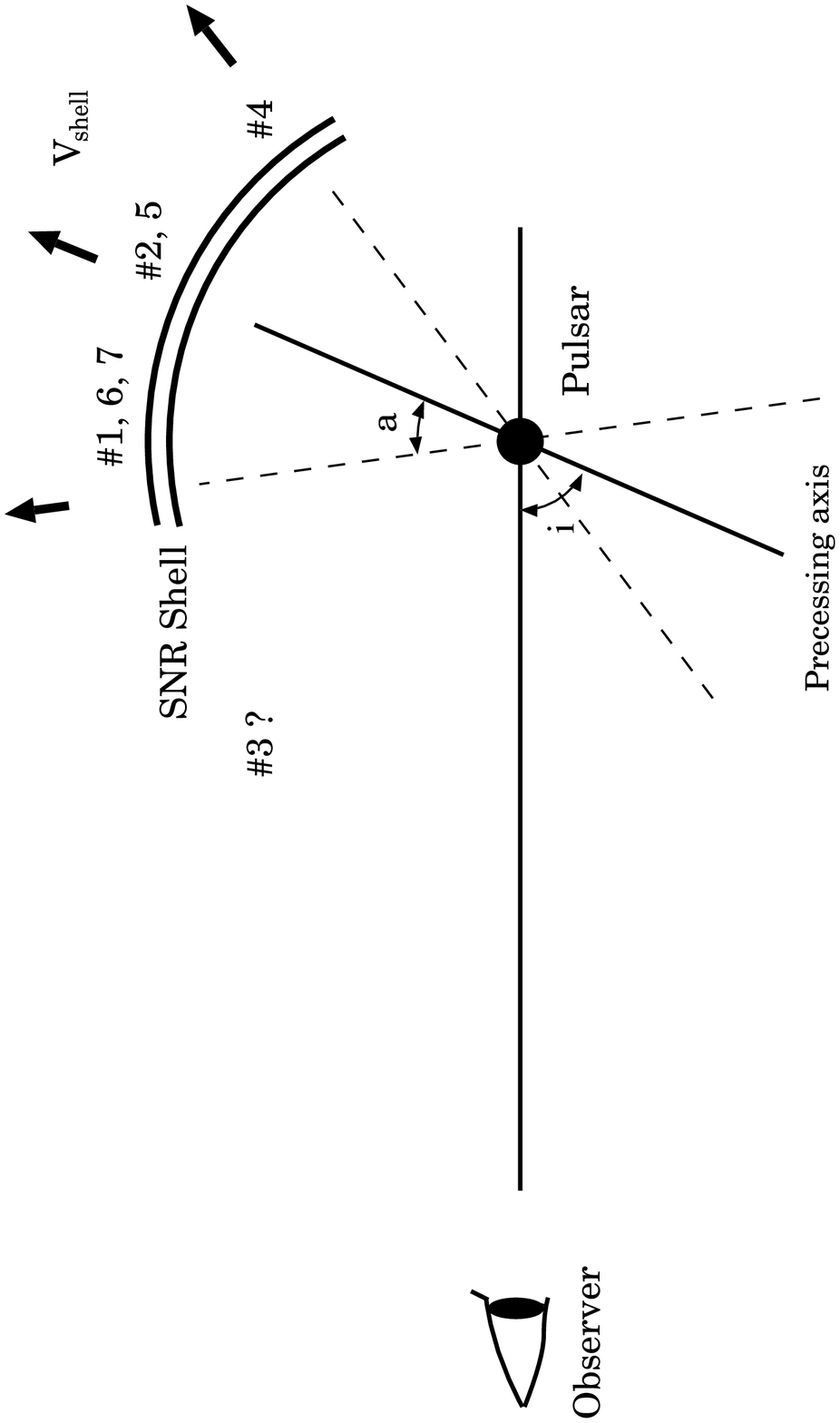}
\caption{Left --- The observed redshift for the each plasma cloud. Right ---
 The schematic image of the geometry of the pulsar and RCW89.  The
 errors are 1 $\sigma$.}
\label{redshift}
\end{figure}

The fits to the thermal emission component of the knots required
non-zero velocity of the emitting gas to account for slight shifts of
the fitted line centroids.  We can constrain the geometry and the
velocity of the ejecta (RCW89) using the velocity of the plasma clouds.
Figure \ref{redshift} (left) shows the observed redshift for the each
knot.  Except for the knot 3 which has the poorest photon statistic, the
redshifts lie on a smooth curve from $0.9\times$ 10$^{-3}$ to $2.9\times
10^{-3}$, which correspond to the velocities of $240\sim 860$ km
s$^{-1}$ along the line of sight.  Now we can consider the geometry as
shown in Figure \ref{redshift} (right).  The SNR system is approaching
us at $\sim70$ km s$^{-1}$ due to the rotation of the Galaxy.  Using the
inclination angle {\itshape i} between the pulsar spin axis and the line
of sight and the half cone angle {\itshape a} of the precession of the
jet, the apparent velocity $v_{\mbox{{\footnotesize app}}}$ is described
as
\begin{eqnarray}
&v_{\mbox{{\footnotesize app max}}} = 
v_{\mbox{{\footnotesize shell}}} \cdot \cos (i - a) -70^{\circ} \quad (\mbox{ km s}^{-1}),& \\
&v_{\mbox{{\footnotesize app min}}} = 
v_{\mbox{{\footnotesize shell}}} \cdot \cos (i + a) -70^{\circ} \quad (\mbox{ km s}^{-1}),&
\label{velocity}
\end{eqnarray}
where $v_{\mbox{{\footnotesize shell}}}$ is the velocity of the shell
boarding RCW89.  Dividing eq. 14 with eq. 15 to cancel
$v_{\mbox{{\footnotesize shell}}}$ yields the relation of $i$ and $a$ as.
\begin{equation}
 \frac{\cos (i-a)}{\cos (i+a)}
=\frac{v_{\mbox{{\footnotesize app max}}} +70^{\circ}}{v_{\mbox{{\footnotesize app min}}} +70^{\circ}}
\end{equation}
The observed X-ray image (Figure \ref{X-ray image}) indicates that the open
angle $a$ is smaller than $20^{\circ}$.  Thus the inclination angle is constrained as
$i>50^{\circ}$, which is consistent with the estimation $i>70^{\circ}$ by
Brazier \& Becker. (1997).  If we know $i$ and $a$, we can calculate the
velocity of the shell.  Using equation 14 and 15 give
\begin{equation}
 v_{\mbox{{\footnotesize shell}}}=
\frac{v_{\mbox{{\footnotesize app max}}}-v_{\mbox{{\footnotesize app min}}}}{2 \sin i \sin a}.
\end{equation}

For the inclination angle $i>50^{\circ}$
($a<20^{\circ}$), the expanding velocity of the shell is obtained as
$v_{\mbox{{\footnotesize shell}}}>1100$ km s$^{-1}$ (For
$v_{\mbox{{\footnotesize shell}}} \sim 4000$ km s$^{-1}$, $i \sim
80^{\circ}$ and $ a \sim 5^{\circ} $ are required ).

\section{Summary}
\begin{figure}[h]
\begin{center}
\includegraphics[height=8.cm]{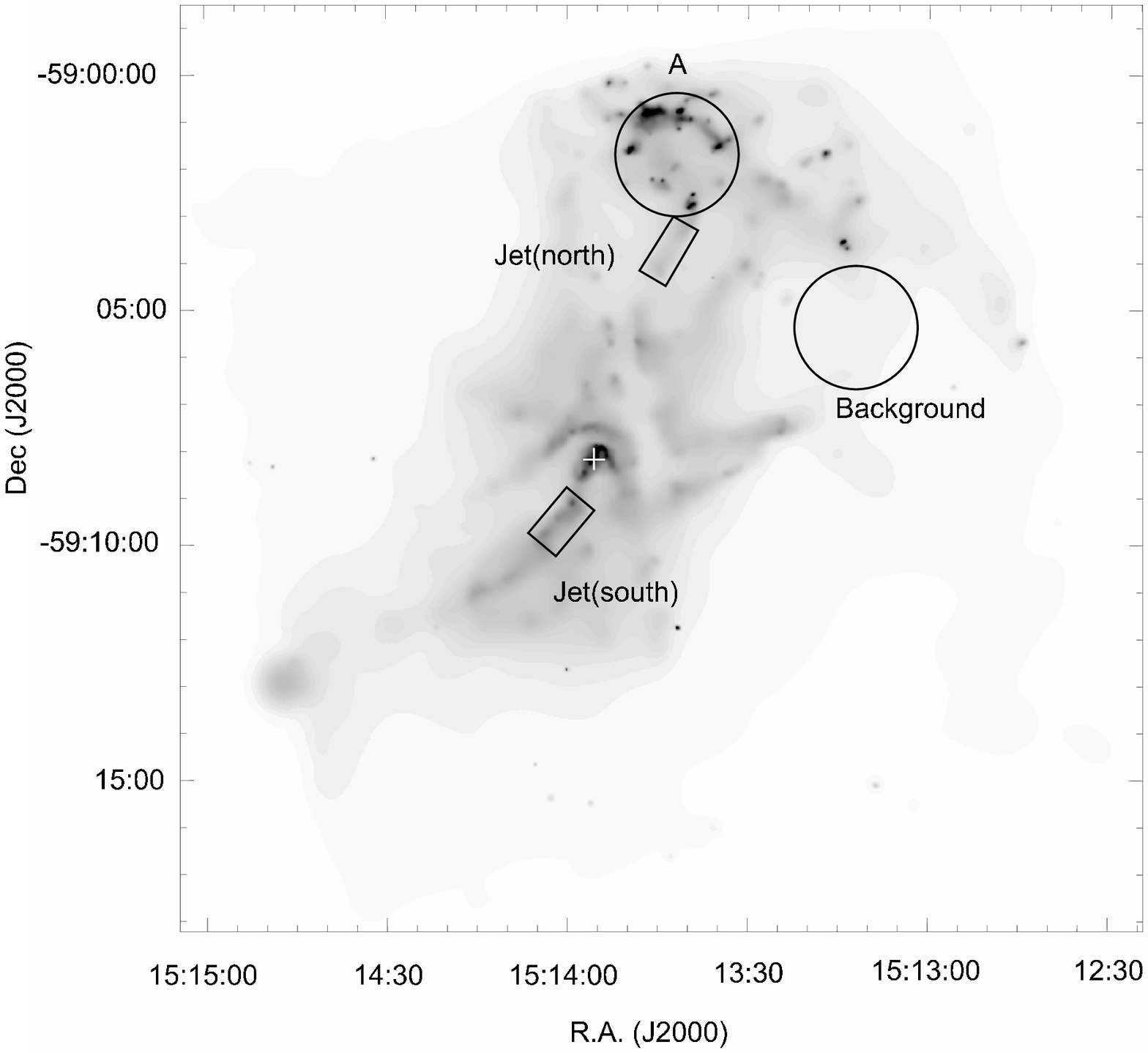}
\includegraphics[height=7.cm]{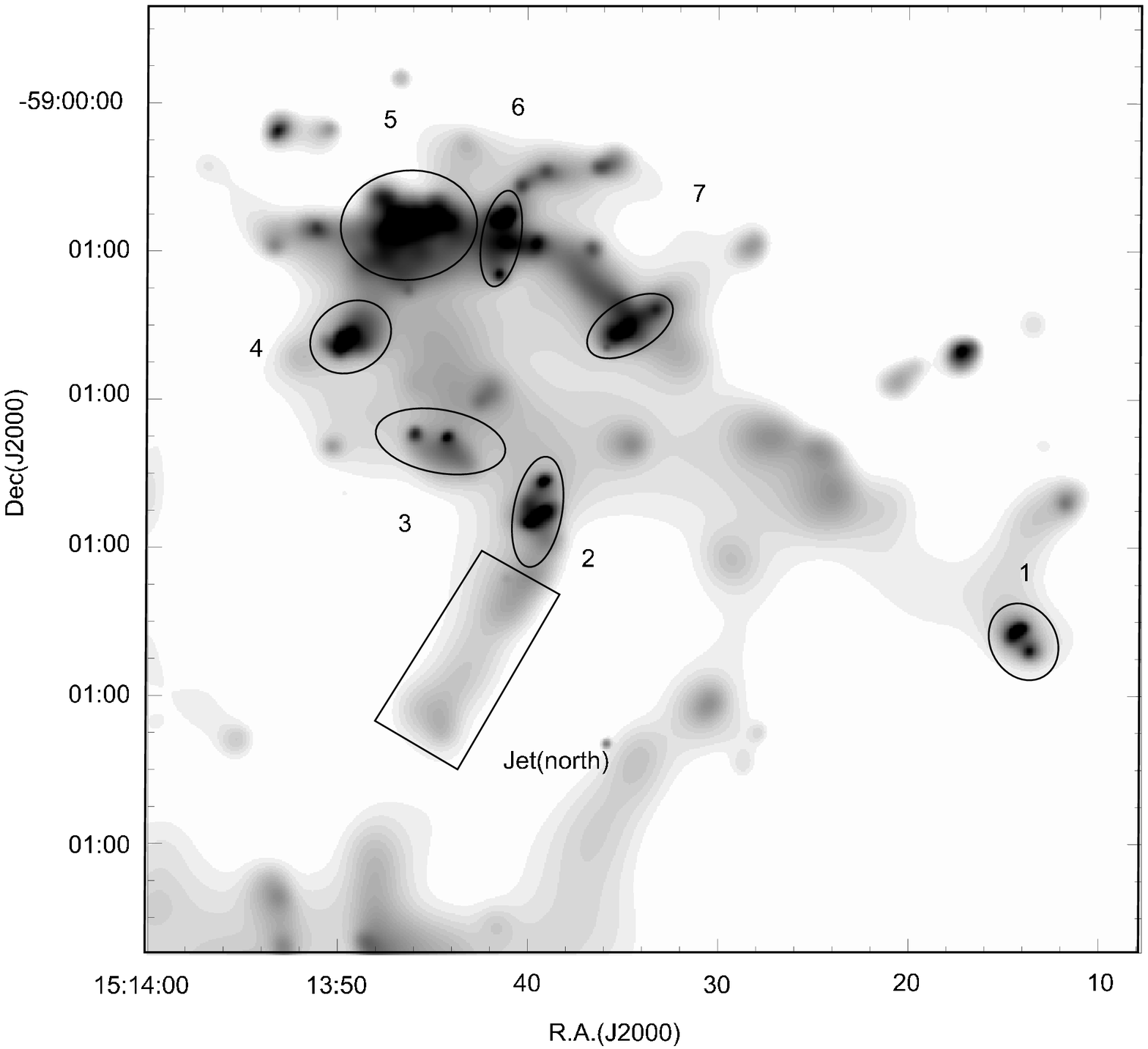} 
\caption{ (left) An overall
X-ray image of RCW89 and PSR B1509--58 in the energy range 0.4$-$8.0
keV.  The pulsar position is marked with a plus sign.  (right) A zoom up
of the interacting region in RCW89 The ellipses and rectangular regions
were selected for succeeding spectral analysis.}  \label{X-ray image}
\end{center}
\end{figure}

\begin{figure}[h]
\begin{center}
\includegraphics[height=8.5cm,angle=-90]{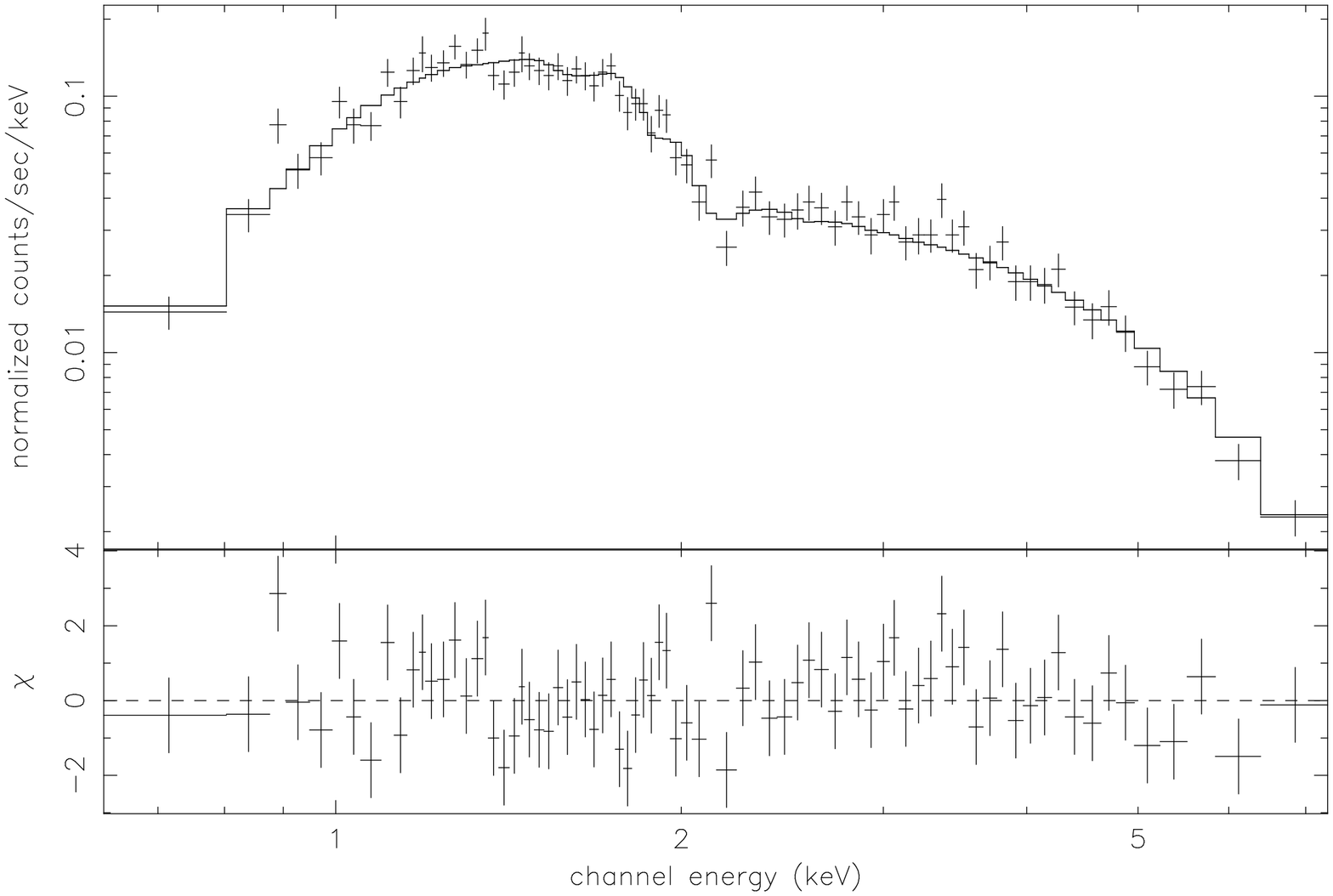}
\includegraphics[height=8.5cm,angle=-90]{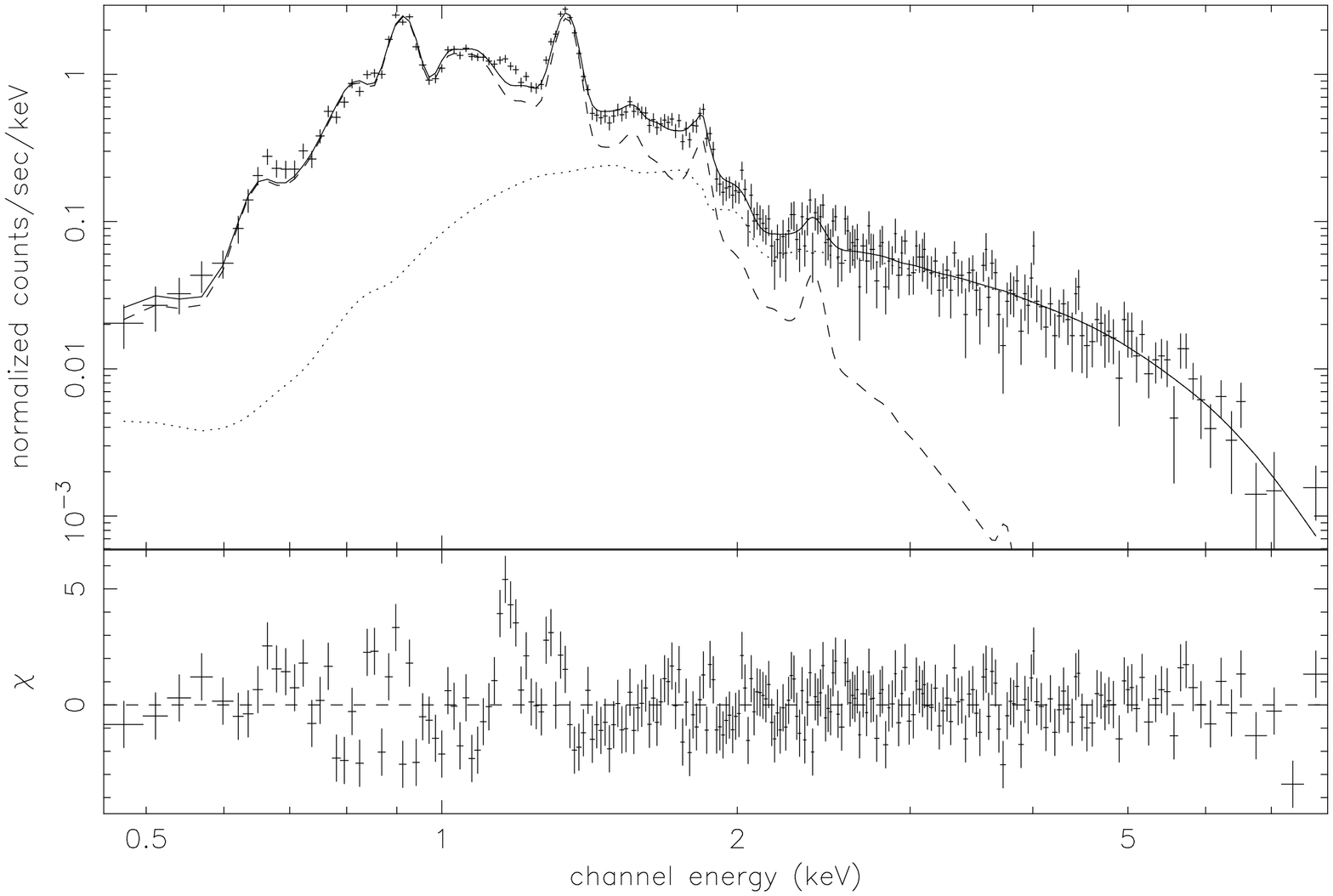}
\caption{ Left: Power law fit to the jet region(labeled in
Fig. \ref{X-ray image}) --- Right: Model fit to the brightest region in
RCW89 labeled ``A'' in Fig. \ref{X-ray image}.  We used a
non-equilibrium ionization model, ``VNEI'' in XSPEC, for the softer
thermal component (dashed line) and a power law model for the hard
component(doted line).}  \label{spectra2}
\end{center}
\end{figure}

We presented a {\itshape Chandra} study of the H {\rm \footnotesize II}
region RCW89.  The observed X-ray image (Figure \ref{X-ray image})
revealed the complicated clump structure distributed along a
``horse-shoe'' shape.  To investigate each plasma cloud, we selected the
7 brightest knots marked in Figure \ref{X-ray image} and performed
spectral analysis for each knot individually.  In the model fitting we
assumed a NEI model and power law function.

The resulting temperature tend to increases along the ``horse-shoe'' in
the clockwise direction, while the ionization time scale $\tau$
decreases.  The gradation of temperature and $\tau$ indicate that the
knots on the north edge of RCW89 are freshly heated hot plasma and are
still in the state of non-equilibrium ionization, while the plasma in
the south region are older and more close to ionization equilibrium.
This new finding is consistent with the picture that RCW89 nebula is
heated by the pulsar jet sequentially.

To confirm the hypotheses, we examined the energy budget for RCW89.  The
observed emission measure and volume yielded stored thermal energy in
the knots as $E_{th}=2.6 \times 10^{47}$ ergs.  On the other hand, the
pulsar jet is supplying relativistic electrons and magnetic energy of
$3.2\times 10^{35}$ ergs s$^{-1}$ into RCW89 assuming a condition of
equipartition and a bulk velocity of the jet of $v_{\mbox{{\tiny
BLK}}}=0.5c$.  The jet requires $\sim1400$ yr to heat the knots, which is
consistent with the pulsar characteristic age.

Next we discussed the cooling process to explain the gradation of the
temperature.  The most plausible process is adiabatic expansion.
If the expansion velocity is the sonic velocity in the thermal
plasma, it takes $\sim 70$ yr for cooling down the knot-7 with typical
values of internal energy $E_{th} = 1.3\times 10^{46}$ ergs and radius of
$r=2\times10^{17}$ cm from 0.6 keV to 0.3 keV.  It is sufficient to form
the temperature gradient within the pulsar time scale of 1700 yr.

The spectral fitting provides information on the metal abundance of each
knot.  The Fe-poor spectra indicate that the matter consisting RCW89
originated in a Type-II supernova.  It is consistent with the existence
of the pulsar.

The obtained redshifts of the each knots vary with their locations.  We
estimated the geometry of the pulsar and the velocity of the shell of
the SNR using the Doppler shifts of the spectra and the X-ray image.
With the precession half cone angle of $ a< 20^{\circ}$, the inclination
angle is constrained as $i > 50^{\circ}$, and the velocity of the shell
as $v_{\mbox{{\footnotesize shell}}} > 1100$ km s$^{-1}$.



\begin{deluxetable}{l c c c}
\tablenum{1}
\tablecolumns{3}
\tablewidth{15cm}
\tablecaption{Spectral fits to region ``A'' and ``jets'' in 
Figure \ref{X-ray image}. }
\tablehead{Parameters & region-A & jet (north) & jet (south)}
\startdata

Column density (10$^{22}$cm$^{-2}$) 
& $1.18\pm0.01$ 
& $0.86\pm0.09$ 
& $1.02\pm0.08$ \\

\sidehead{{\itshape Thermal component}}
Temperature   (keV) 
& $0.38 \pm 0.01$ 
&
&\\

$n_e t$\tablenotemark{a} ($10^{11}$ s cm$^{-3}$) 
& $0.71\pm0.05$
& 
&\\

$E\!M$\tablenotemark{b}  ($10^{57}$cm$^{-3}$) 
& $3.42\pm 0.03$ 
&
& \\

\sidehead{Abundance (solar abundance)}
Ne & $1.00 \pm 0.03$ & &\\
Mg & $0.92 \pm 0.02$ & &\\
Si & $0.41 \pm 0.04$ & &\\
Fe & $0.32 \pm 0.02$ & &\\

\tableline

\sidehead{{\itshape Non-thermal component}}

Photon index  
& $2.48\pm0.05$ 
& $2.07\pm0.11$
& $1.66\pm0.09$\\

Flux$_{0.5 - 5.0 \mbox{{\tiny keV}}}$  ($10^{-12}$ ergs s$^{-1}$ cm$^{-2}$) 
& $8.67 \pm 0.05$ 
& $1.26 \pm 0.09$
& $4.16 \pm 0.26$ \\

\tableline

$\chi_{\nu}^2 \; (\nu)$ 
& 1.727 (227) 
& 0.981 (70)
& 0.493 (172)\\
\enddata 

\tablecomments{Errors are with 1 $\sigma$ confidence.}
\tablenotetext{a}{Ionization timescale of the non-equilibrium ionization
plasma.}  
\tablenotetext{b}{Emission measure $ \int n_e n_H dV 
\left(d_L/5.2 \mbox{kpc}\right)^2$ ( $d_L$ is the distance to RCW89).  }
\end{deluxetable}

\begin{deluxetable}{lcccccccccrl}
\rotate
\tabletypesize{\footnotesize}
\tablecolumns{11}
\tablewidth{0pc}
\tablecaption{Results from spectra fitting}
\tablenum{2}
\tablehead{
 \colhead{Region \tablenotemark{a}}
 &\colhead{$kT$}
 &\colhead{$n_e t$ \tablenotemark{b}}
 &\colhead{$E\!M$ \tablenotemark{c}}
 &\colhead{$F_{\mbox{{\footnotesize th}}}$ \tablenotemark{d}}
 &\colhead{$F_{\mbox{{\footnotesize PL}}}$ \tablenotemark{d}}
 &\multicolumn{4}{c}{Abundance}
 &\colhead{Redshift}
 &\colhead{$\chi^2_{\nu}\quad (\nu)$}
\\
  \colhead{}
 &\colhead{(keV)}
 &\colhead{($10^{11}$cm$^{-3}$s)}
 &\colhead{(10$^{56}$cm$^{-3}$)}
 &\colhead{($\times10^{-11}$)}
 &\colhead{($\times10^{-13}$)}
 &\colhead{Ne}
 &\colhead{Mg}
 &\colhead{Si}
 &\colhead{Fe}
 &\colhead{($\times 10^{-3}$)}
 &\colhead{}
 }
 \startdata

 1.....
 &$ 0.21_{- 0.02 }^{+ 0.01 }$
 &$>11(24.2)$
 &$ 3.94_{- 2.28 }^{+ 1.17 }$
 &$ 1.93_{- 0.57 }^{+ 1.14 }$
 &$ 2.33_{- 0.34 }^{+ 0.30 }$
 &$ 1.64_{- 0.24 }^{+ 0.24 }$
 &$ 2.46_{- 0.36 }^{+ 0.68 }$
 &$ 1.02_{- 1.02 }^{+ 1.04 }$
 &$<0.17(0.00)$
 &$ 0.83_{- 0.19 }^{+ 0.27 }$
 &1.387 (39)
 \\
 2.....
 &$ 0.26_{- 0.01 }^{+ 0.02 }$
 &$>11(28.8)$
 &$ 2.30_{- 0.52 }^{+ 0.64 }$
 &$ 1.29_{- 0.20 }^{+ 0.31}$
 &$ 4.17_{- 0.42 }^{+ 0.42 }$
 &$ 1.94_{- 0.23 }^{+ 0.22 }$
 &$ 2.99_{- 0.30 }^{+ 0.61 }$
 &$<0.90(0.40)$
 &$ 0.62_{- 0.13 }^{+ 0.12 }$
 &$ 2.74_{- 0.81 }^{+ 0.13 }$
 & 1.150 (62)\\
 3.....
 &$ 0.20_{- 0.03 }^{+ 0.04 }$
 &$ 1.46_{- 0.68 }^{+ 1.59 }$
 &$11.56_{- 5.47 }^{+ 9.14 }$
 &$ 6.80_{- 1.29 }^{+ 3.24 }$
 &$ 5.33_{- 0.41 }^{+ 0.41 }$
 &$ 0.54_{- 0.15 }^{+ 0.14 }$
 &$ 1.02_{- 0.19 }^{+ 0.87 }$
 &$ 1.16_{- 0.56 }^{+ 0.53 }$
 &$ 0.14_{- 0.13 }^{+ 0.34 }$
 &$-2.50_{- 2.62 }^{+ 3.05 }$
 & 1.083 (50)\\
 4.....
 &$ 0.41_{- 0.01 }^{+ 0.02 }$
 &$ 1.07_{- 0.13 }^{+ 0.16 }$
 &$ 1.43_{- 0.04 }^{+ 0.16 }$
 &$ 1.90_{- 0.32 }^{+ 0.50 }$
 &$ 2.01_{- 0.42 }^{+ 0.34 }$
 &$ 1.52_{- 0.14 }^{+ 0.13 }$
 &$ 1.46_{- 0.11 }^{+ 0.13 }$
 &$ 0.35_{- 0.12 }^{+ 0.13 }$
 &$ 0.30_{- 0.06 }^{+ 0.07 }$
 &$ 2.88_{- 0.15 }^{+ 1.48 }$
 &1.682 (61)\\
 5.....
 &$ 0.40_{- 0.01 }^{+ 0.01 }$
 &$ 0.70_{- 0.05 }^{+ 0.07 }$
 &$ 8.22_{- 0.29 }^{+ 0.28 }$
 &$ 9.89_{- 0.48 }^{+ 0.53 }$
 &$ 3.11_{- 0.59 }^{+ 0.59 }$
 &$ 1.18_{- 0.05 }^{+ 0.05 }$
 &$ 0.93_{- 0.04 }^{+ 0.04 }$
 &$ 0.50_{- 0.04 }^{+ 0.06 }$
 &$ 0.41_{- 0.04 }^{+ 0.02 }$
 &$ 2.18_{- 0.29 }^{+ 0.24 }$
 &1.674 (104)\\
 6.....
 &$ 0.52_{- 0.02 }^{+ 0.02 }$
 &$ 0.46_{- 0.05 }^{+ 0.06 }$
 &$ 1.17_{- 0.12 }^{+ 0.06 }$
 &$ 2.94_{- 0.38 }^{+ 0.48 }$
 &$ 2.96_{- 0.46 }^{+ 0.47 }$
 &$ 2.16_{- 0.14 }^{+ 0.15 }$
 &$ 1.77_{- 0.11 }^{+ 0.10 }$
 &$ 0.54_{- 0.11 }^{+ 0.11 }$
 &$ 0.44_{- 0.06 }^{+ 0.07 }$
 &$ 0.95_{- 0.21 }^{+ 0.10 }$
 &1.528 (77)\\
 7.....
 &$ 0.67_{- 0.08 }^{+ 0.08 }$
 &$ 0.23_{- 0.05 }^{+ 0.08 }$
 &$ 0.53_{- 0.14 }^{+ 0.13 }$
 &$ 1.69_{- 0.41 }^{+ 0.42 }$
 &$ 3.45_{- 0.80 }^{+ 0.64 }$
 &$ 2.16_{- 0.29 }^{+ 0.48 }$
 &$ 1.53_{- 0.23 }^{+ 0.11 }$
 &$ 0.38_{- 0.18 }^{+ 0.22 }$
 &$ 0.12_{- 0.07 }^{+ 0.12 }$
 &$ 0.80_{- 0.11 }^{+ 0.25 }$
 &1.335 (64)\\
 \enddata

\tablecomments{Errors are with 1 $\sigma$ confidence.  The 1 $\sigma$
 limits (with the best-fit values in the parentheses) are presented
 where the best-fit parameters are unconstrained.}

\tablenotetext{a}{Regions as marked in Fig. \ref{X-ray image}.}

\tablenotetext{b}{Ionization timescale for the non-equilibrium
 ionization plasma.}

\tablenotetext{c}{Emission measure $\int n_e n_H dV \left(d_L/5.2\mbox{kpc}\right)^2$  
($d_L$ is the distance toward RCW89.)}

\tablenotetext{d}{$F_{\mbox{{\footnotesize th}}}$ and
 $F_{\mbox{{\footnotesize PL}}}$ are the unabsorbed energy flux in 0.5
 $-$ 5.0 keV for the thermal component and for the power-law function,
 respectively.  They are expressed in a unit of ergs cm$^{-2}$ s$^{-1}$.
 }

\end{deluxetable}

\begin{deluxetable}{lccccc}
\tablecolumns{6}
\tablewidth{15cm}
\tablecaption{Estimated plasma parameters assuming $d_L = 5.2$ kpc}
\tablenum{3}
\tablehead{
\colhead{Region \tablenotemark{a}}
 &\colhead{$l_a \times l_b$ \tablenotemark{b}}
 &\colhead{$ V$  \tablenotemark{c}}
 &\colhead{$ n_e $ }
 &\colhead{$ E_{\mbox{{\footnotesize thermal}}}$ \tablenotemark{d}}
 &\colhead{$ t_{\mbox{{\footnotesize plasma}}}$ \tablenotemark{e}}
\\
\colhead{}
&\colhead{(arcsec)}
&\colhead{(cm$^{3}$)}
&\colhead{($f^{-1/2}$ cm$^{-3}$)}
&\colhead{($f^{1/2}$ ergs)}
&\colhead{($f^{1/2}$ yr)}
 }
\startdata
1....... 
 &$2.0 \times 1.7 $
 &
 &
 &
 &
\\
 &$ 1.5 \times 1.0 $
 &$ 1.5 \times 10^{ 52 }$ 
 &$ 510 _{ -180 }^{ +70}$
 &$  7.7_{  -3.2}^{ +1.5 } \times 10^{ 45 }$ 
 &$>59$(150)
\\
2....... 
 &$8.0 \times 2.5 $
 &
 &
 &
 &
\\
 &$2.0 \times 2.0$
 & $ 1.2\times 10^{53}$ 
 & $140_{  -20}^{ +20 }$
 & $2.1_{  -0.3}^{ +0.4 }  \times 10^{46}$ 
 & $>220$(660)\\
3....... 
 &$2.2 \times 2.0$
 &
 &
 &
 &
\\
 & $2.7 \times 2.7$
 & $5.7 \times 10^{52}$ 
 & $450_{  -120}^{ +150 } $
 & $2.5_{  -0.9}^{ +1.5 } \times 10^{46}$ 
 & $10_{   -6  }^{ +20  }$ \\
4....... 
 & $5.0 \times 2.4$
 & $5.8 \times 10^{52}$ 
 & $160_{  -3}^{ +9 }$
 & $1.8_{  -0.1}^{ +0.2 } \times 10^{46}$ 
 & $22_{  -4}^{ +4 }$ \\
5....... 
 & $15.7 \times 4.9$
 & $7.5 \times 10^{53}$ 
 & $100_{  -2}^{ + 2}$ 
 & $1.5_{  -0.6}^{ +0.6 } \times 10^{47}$ 
 & $21_{   -2}^{ +3 } $\\
6....... 
 & $3.4 \times 2.0$
 &
 &
 &
 &
\\
 & $11.3 \times 2.7 $
 &
 &
 &
 &
\\
 & $1.7 \times 1.7$
 & $2.0 \times 10^{53}$ 
 & $76_{  -4}^{  +2 }$
 & $3.8_{  -0.3}^{ +0.2 } \times 10^{46}$ 
 & $19 _{ -2 }^{ +4 }$
 \\
7....... 
 &$3.8 \times 2.0$
 & $3.0 \times 10^{52}$ 
 & $130_{  -18}^{ +15 } $ 
 & $1.3_{  -0.3}^{ +0.3 } \times 10^{46}$ 
 & $5_{  -2}^{ +3 }$
\\
 \enddata
\tablecomments{Uncertainties are all at 1 $\sigma$ confidence.  The 1
 $\sigma$ limits (with the best values in the parentheses) are presented
 where the best-fit parameters are unconstrained.  There still remains
 uncertainty $f$, which is the filling factor or the uncertainty of the
 emitting volume.}
\tablenotetext{a}{Regions as marked in Fig. \ref{X-ray image}}
\tablenotetext{b}{Size of the plasma cloud.  The axis length
 ($2l_a, 2l_b$) are defined as the FWHM of the surface brightness. For
 the region 1, 2, 3, and 6, we measured the size of each distinctive cloud.  }
\tablenotetext{c}{Total volume of the plasma clouds in the region.}
\tablenotetext{d}{Internal thermal energy contained in the knots.}
\tablenotetext{e}{Plasma age obtained from the ionization parameter and
 electron number density ($n_e t / n_e$).}
\end{deluxetable}

\begin{thebibliography}{}
\bibitem{Brazier97} Brazier, K. T. S., \& Becker,
			    W. 1997, MNRAS, 284, 335
\bibitem{}Brinkmann, W., Aschenbach, B., \& Langmeier, A. 1985, Nature,
			    313, 662
\bibitem{}Caswell, J. L., Milne, D. K., \& Wellington, K. J. 1981 MNRAS,
	195, 89
\bibitem[Gaensler B. M. 2002]{Gaensler02} Gaensler, B. M., Arons, J., Kaspi,
			    V. M., Pivovaroff, M. J., Kawai, N., \& Tamura,
			    K. 2002, \apj, 569, 878
\bibitem[Gaensler B. M.1999]{Gaensler99} Gaensler, B. M., Brazier,
			    K. T. S., Manchester, R. N., Johnston, S.,
			    \& Green, A. J. 1999, MNRAS, 305, 724
\bibitem[Gaensler B. M.1998]{Gaensler98} Gaensler, B. M., Brazier,
			    K. T. S., Manchester, R. N., Johnston, S.,
			    \& Green, A. J. 1998, MmSAI, 69, 877
\bibitem{} Helfand, D. J., Gotthelf, E. V., \& Halpern, J. P. 2001, \apj,
	556, 380
\bibitem{} Hester, J. J. et al. 1995, \apj, 448, 240
\bibitem{} Hester, J. J. et al. 2002, \apj, 577, L49
\bibitem{} Kaspi, V. M., et al. 1994, \apj, 422, L83
\bibitem{} Manchester, R. N., Tuohy, I. R., \& D'Amico, N. 1982, \apj,
	262, L31
\bibitem{} Manchester, R. N., 1987, A\&A, 171, 205
\bibitem{} Pavlov, G. G., Teter, M. A., Kargaltsev, O., \& Sanwal,
	D. 2002, \apj, 591, 1157 
\bibitem{} Rodgers A. W., Campbell C. T., Whiteoak J.B., 1960 MNRAS,
				    121, 103
\bibitem[Sako T.2000]{Sako00} Sako, T., et al. 2000, \apj, 537, 422
\bibitem{} Seward, F. D., \& Harnden, F. R. Jr., 1982, \apj, 256, L45
\bibitem[Tamura K.1996]{Tamura96} Tamura, K., Kawai, N., Yoshida, A., \&
			    Brinkmann, W. 1996, PASJ, 48, L33
\bibitem[Trussoni.E]{Trussoni96} Trussoni, E., Massaglia, S., Caucino,
			    S., Brinkmann, W., \& Aschenbach, B. 1996,
			    A\&A, 306, 581
\bibitem{} Weisskopf et al. 2000, \apj, 536, L81
\bibitem{} Whiteoak, J. B. Z., Green, A. J. 1996, A\&AS, 118, 329
\end{thebibliography}
\end{document}